\documentclass[a4paper,fleqn]{cas-sc}
\usepackage[numbers]{natbib}
\usepackage{hyperref}
\usepackage[dvipsnames]{xcolor}

\def\tsc#1{\csdef{#1}{\textsc{\lowercase{#1}}\xspace}}
\tsc{WGM}
\tsc{QE}
\tsc{EP}
\tsc{PMS}
\tsc{BEC}
\tsc{DE}

\begin{document}
\let\WriteBookmarks\relax
\def\floatpagepagefraction{1}
\def\textpagefraction{.001}

\shorttitle{Independence role in the generalized Sznajd model}

\shortauthors{Azhari et.al}

\title[mode = title]{Independence role in the generalized Sznajd model}    
\author[usu1,usu2]{Azhari}[orcid= 0000-0003-2178-330X]
\ead{azhari@usu.ac.id}
\cormark[1]
\cortext[cor1]{Corresponding author}

\author[brin]{Roni Muslim}[orcid= 0000-0001-6925-5923]
\ead{roni.muslim@brin.go.id}

\author[brin,uty]{Didi Ahmad Mulya}[orcid= 0009-0003-5061-6217]
\ead{didiahmadmulya1@gmail.com}

\author[inst6]{Heni Indrayani}[orcid= 0000-0003-3659-1191]
\ead{heni.indrayani@dsn.dinus.ac.id}

\author[inst5]{Cakra Adipura Wicaksana}[orcid= 0009-0007-0515-3478]
\ead{cakraadipura@untirta.ac.id}

\author[inst4]{Akbar Rizki}[orcid= 0000-0002-5187-9356]
\ead{akbar.ritzki@apps.ipb.ac.id}

\affiliation[usu1]{
    organization={Department of Physics, Universitas Sumatera Utara},
    city={Medan},
    postcode={20155},
    country={Indonesia}
}

\affiliation[usu2]{
    organization={Integrated Research Laboratory, Universitas Sumatera Utara},
    city={Medan},
    postcode={20155},
    country={Indonesia}
}

\affiliation[brin]{
    organization={Research Center for Quantum Physics,  National Research and Innovation Agency (BRIN)},
    city={South Tangerang},
    postcode={15314},
    country={Indonesia}
}

\affiliation[uty]{
    organization={Department of Industrial Engineering,  University of Technology Yogyakarta},
    city={Yogyakarta},
    postcode={55285},
    country={Indonesia}
}

\affiliation[inst6]{
    organization={Department of Communication Science, Universitas Dian Nuswantoro},
    city={Semarang},
    postcode={50277},
    country={Indonesia}
}

\affiliation[inst5]{
    organization={Department of Electrical Engineering, Universitas Sultan Ageng Tirtayasa},
    city={Serang},
    postcode={42435},
    country={Indonesia}
}

\affiliation[inst4]{
    organization={Department of Statistics, IPB University},
    city={Bogor},
    postcode={16680},
    country={Indonesia}
}

\begin{abstract}
The Sznajd model is one of sociophysics's well-known opinion dynamics models. Based on social validation, it has found application in diverse social systems and remains an intriguing subject of study, particularly in scenarios where interacting agents deviate from prevailing norms. This paper investigates the generalized Sznajd model, featuring independent agents on a complete graph and a two-dimensional square lattice. Agents in the network act independently with a probability \(p\), signifying a change in their opinion or state without external influence. This model defines a paired agent size \(r\), influencing a neighboring agent size \(n\) to adopt their opinion. This study incorporates analytical and numerical approaches, especially on the complete graph. Our results show that the macroscopic state of the system remains unaffected by the neighbor size \(n\) but is contingent solely on the number of paired agents \(r\).
Additionally, the time required to reach a stationary state is inversely proportional to the number of neighboring agents \(n\). For the two-dimensional square lattice, two critical points \(p = p_c\) emerge based on the configuration of agents. The results indicate that the universality class of the model on the complete graph aligns with the mean-field Ising universality class. Furthermore, the universality class of the model on the two-dimensional square lattice, featuring two distinct configurations, is identical and falls within the two-dimensional Ising universality class.
\end{abstract}





\maketitle

\section{Introduction}
During the past decade, science has progressed rapidly, with numerous disciplines forming connections. Physicists who specialize in statistical physics and nonlinear phenomena, for instance, have tried implementing relevant concepts to understand social and political phenomena~\cite{serge2016sociophysics,sen2014sociophysics,javarone2014network,castellano2009statistical,azhari2023external}. This field is commonly known as sociophysics, interdisciplinary science that explores various socio-political phenomena using the principles and concepts of statistical physics. One of the most popular topics in sociophysics is opinion dynamics model~\cite{serge2016sociophysics, sen2014sociophysics,castellano2009statistical, stauffer2009encyclopedia}, which involves modeling the interactions of agents interconnected within a network topology. To analyze and predict various social phenomena, such as transition states, hysteresis, critical mass, and many more, physicists have attempted to correlate micro- and macro-scale physical system phenomena with social structures~\cite{myer2013}.

Since developing opinion dynamics models aims to explain various social phenomena, creating realistic models has been one of the biggest challenges for scientists. Several opinion dynamics models have been proposed in both discrete and continuous forms. Examples include the Sznajd model~\cite{sznajd2000opinion}, the voter model~\cite{liggett1985interacting}, the majority rule model~\cite{mobilia2003majority, galam2002minority, krapivsky2003dynamics}, the Biswas-Sen model~\cite{biswas2009model, biswas2023social}, and the Galam model~\cite{galam2008sociophysics}. These models have emerged from physicists studying similar thermodynamics and statistical physics correlations. Most models exhibit a ferromagnetic-like quality, ensuring the system remains homogeneous, meaning all members eventually maintain the same opinion. In sociological research~\cite{nail2000proposal}, this ferromagnetic characteristic represents conformity behavior. However, when compared to real social situations, these models fail to reflect reality accurately. To make the models more realistic, physicists have introduced several social parameters, such as nonconformity~\cite{nyczka2013anticonformity}, inflexibility~\cite{galam2007role}, contrarianism~\cite{galam2004contrarian}, and fanaticism~\cite{mobilia2003does}, to create more complex dynamics that better correlate with various social phenomena.

Based on the objectives of dynamics modeling, it is intriguing to consider destructive social behaviors described in social psychology, such as independence and anticonformity behaviors~\cite{nail2000proposal, willis1963two, willis1965conformity, macdonald2004expanding, nail2011development}. As Milgram stated~\cite{milgram1963behavioral}, "\textit{Independent behavior refers to the ability to resist pressures to conform to a majority or resist pressures to obey the orders given by an authority figure}." In other words, an independent agent acts without being influenced by the group. This behavior disrupts social cohesion by acting outside the control of the majority and plays a significant role in social dynamics. Anticonformity is a behavior that rejects adopting the majority opinion. Anticonformity and independence differ in their relationship to the group's influence: Anticonformity actively evaluates and opposes the group's opinion, whereas independence ignores the group's opinion. Particularly for the behavior of independence, several previous studies have examined the impact of independence in various opinion dynamics models, such as the \( q \)-voter model and the majority-rule model \cite{chmiel2015phase,crokidakis2015inflexibility,vieira2016consequences,vieira2016phase,radosz2017q,abramiuk2020generalized,doniec2022consensus,oestereich2023phase}. Under different scenarios, these studies have shown significant effects on the macroscopic state of the system, including the occurrence of order-disorder phase transitions.

The implementation of independent social behavior in the Sznajd model with various scenarios and network topologies can be seen in Refs.~\cite{sznajd2011phase, karan2017modeling, muslim2022opinion}. The authors define the Sznajd model on complete graphs, including one- and two-dimensional square lattices. They also introduced the flexibility parameter, which describes how likely an agent is to change its opinion. Based on the results, the models on the complete graph and the two-dimensional square lattice undergo a continuous phase transition characterized by a diminishing critical point as the flexibility parameter value increases. Conversely, it is noteworthy that no discernible phase transition was observed in the model defined on the one-dimensional lattice \cite{sznajd2011phase}.
In Ref.~\cite{karan2017modeling}, the authors studied the Sznajd model using the master equation to analyze the associated dynamics. Both analytical approaches and numerical simulations have shown that the convergence of magnetization depends on the initial influencer distribution. A recent study examined the Sznajd model defined on a complete graph with two different agent configurations, namely three-against-one and two-against-two configurations \cite{muslim2022opinion}. Independent agents and flexibility factors are also introduced as control parameters for the occurrence of the order-disorder phase transition. Based on the analytically and numerically obtained results, the model undergoes a continuous phase transition for both configurations, with the critical point depending on the flexibility factor. However, the interaction between agents is limited, so in that model \cite{muslim2022opinion} dynamics, information cannot be obtained for other cases, such as how the number of influencers depends on the time to reach equilibrium and other macroscopic phenomena.

This paper discusses a more generalized Sznajd model compared to previous versions, considering influencer agents of size \( r \) and neighboring agents to be persuaded of size \( n \). In specific cases, this model is reduced to the original Sznajd model~\cite{sznajd2000opinion} and the \( q \)-voter model~\cite{castellano2009nonlinear} defined on the complete graph. Like previous studies, this model is defined on the complete graph where all agents are neighbors. Additionally, we consider the model on a two-dimensional square lattice with two different influencer configurations: one where only four paired agents have homogeneous opinions (case one)  can influence their nearest neighbors and another where not only four paired agents have the same opinion, but if any two influencer agents share the same opinion, they can influence their nearest neighbors to adopt their opinion (case two). 

We investigate the independent behavior of the occurrence of order-disorder phase transitions in the system and analyze the universality class of the model on both the complete graph and the two-dimensional square lattice. Our results, obtained analytically and through Monte Carlo simulations, show that the critical point at which the system undergoes an order-disorder phase transition is influenced only by the number of persuaders \( r \), not by the number of persuaded neighbors \( n \). The phase transition is continuous for \( r \leq 5 \) and discontinuous for \( r > 5 \) for all values of \( n \). The number of persuaded neighbors \( n \) only affects the time it takes for the system to reach equilibrium, following the relation \( t_{\text{steady}} \sim 1/n \). The obtained critical exponents indicate that the model on the complete graph shares the same universality class as the Ising mean-field model. For the two-dimensional square lattice, our results indicate that the model undergoes only a continuous phase transition in both cases, with the critical point for case one being higher than for case two. Despite the difference in critical points, both cases exhibit identical critical exponents, suggesting they belong to the same universality class as the two-dimensional Ising model.

\section{Model and methods}
\label{sec.2}
The original Sznajd model posits that two paired agents with the same opinion can influence two of their neighbors to adopt their opinion (social validation). Mathematically, if \( S_i = S_{i+1} \), then \( S_{i-1} = S_i = S_{i+1} = S_{i+2} \). Conversely, if the paired agents have different opinions, their neighbors adopt the opposite opinions alternately; mathematically, if \( S_i \neq S_{i+1} \), then \( S_{i-1} = S_{i+1} \) and \( S_i = S_{i+2} \)~\cite{sznajd2000opinion}. The final state of the original Sznajd model results in either complete consensus (ferromagnetic) or a stalemate situation (antiferromagnetic). From a social perspective, the final state of the Sznajd model is less representative of real social states. To enhance the model's dynamism and richness, we consider introducing a noise parameter, socially destructive behavior, or independence in the social literature~\cite{hofstede2010} and analyze its impact on the system. The model is defined on the complete graph and the two-dimensional square lattice. Independent behavior, which can naturally disrupt social cohesion, introduces more dynamic phenomena into the model, such as the emergence of phase transitions.

\begin{figure}[tb]
    \centering
    \includegraphics[width =0.8\textwidth]{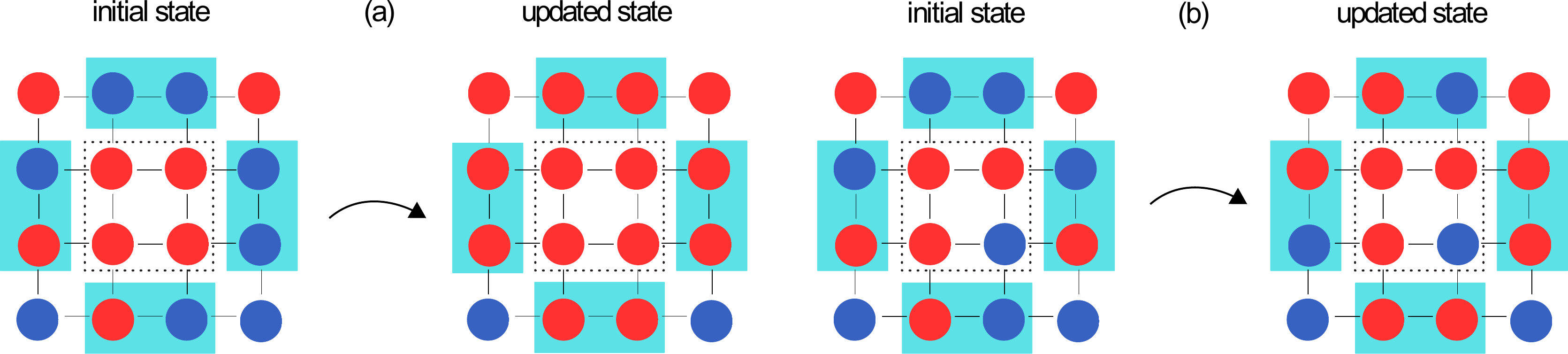}
    \caption{Scheme of the Sznajd model on a two-dimensional square lattice. (a) Only four agents with the same opinion can influence eight neighboring agents to follow them. (b) If the four agents do not share the same opinion, then two agents with the same opinion (in a row or column of the panel) can influence their neighbors to adopt their opinion or adopt alternating opinions. This mechanism is similar to the one-dimensional Sznajd model.}
    \label{fig:2D_square}
\end{figure}

To analyze the macroscopic phenomena in the model, we employ an agent-based model where each agent can have one of two possible opinions, represented by the Ising number \(\sigma_i = \pm 1\). For example, \(+1\) and \(-1\) represent the opinions (states) 'up' and `down,' respectively. This modeling reflects social situations where individuals often have binary choices: pro or contra, yes or no, choose A or B. The agents' opinions are randomly assigned to the nodes of a graph, with the graph links representing social connections. To analyze the macroscopic parameters of the system, we initialize the system in a disordered state, meaning the total population of agents with `up' and `down' opinions is equal and randomly distributed across the network. The algorithm for the model can be described as follows:
\begin{itemize}
    \item The model on the complete graph.
    \begin{enumerate}
        \item We randomly select a group of agents \( r \) to influence their neighbors \( n \). With probability \( p \), the neighbor agents adopt independent behavior. In this stage, there is no opinion change among the neighbor agents. With a probability of \( 1/2 \), each agent from the neighbor agents \( n \) flips their opinion from \( +1 \) to \( -1 \) or vice versa, depending on their initial opinion. The opinion change of the neighbor agents is not influenced by the group of agents \( r \).
        \item Suppose the neighbor agents do not adopt independent behavior. In that case, they will adopt the opinion of the group of agents \( r \) with probability \( 1 - p \), provided the group of agents \( r \) shares the same opinion.
    \end{enumerate}
    \item The model on the two-dimensional square lattice
    \begin{enumerate}
        \item Case 1
        \begin{enumerate}
            \item We randomly select a group of agents \( r \) consisting of four agents (influencers): \( S_{(i,j)}, S_{(i+1,j)}, S_{(i+1,j-1)}, \\ S_{(i,j-1)} \). With probability \( p \), their eight nearest neighbors adopt independent behavior. In this stage, there is no opinion change among the neighbor agents. With a probability of \( 1/2 \), each agent from the eight neighbors flips their opinion from \( +1 \) to \( -1 \) or vice versa, depending on their initial opinion. The opinion change of the neighbor agents is not influenced by the group of agents \( r \).
            \item Suppose the eight neighbor agents do not adopt independent behavior. In that case, they will adopt the opinion of the four influencer agents with probability \( 1 - p \), provided the four agents have the same opinion, as shown in Fig.~\ref{fig:2D_square} (a).
        \end{enumerate}
        \item Case 2
        \begin{enumerate}
            \item We randomly select a group of agents \( r \) consisting of four agents (influencers): \( S_{(i,j)}, S_{(i+1,j)}, S_{(i+1,j-1)}, \\ S_{(i,j-1)} \). With probability \( p \), their eight nearest neighbors adopt independent behavior. In this stage, there is no opinion change among the neighbor agents. With a probability of \( 1/2 \), each agent from the eight neighbors flips their opinion from \( +1 \) to \( -1 \) or vice versa, depending on their initial opinion. The opinion change of the neighbor agents is not influenced by the group of agents \( r \).
            \item Suppose the eight neighboring agents do not adopt independent behavior. In that case, even though the four agents \(r\) do not share the same opinion, the eight nearest neighboring agents still have the potential to change their opinions, following the one-dimensional Sznajd model. Two pairs of agents along the row and column directions can influence their two nearest neighbors. If both pairs of agents in the row and column directions agree (i.e., have the same opinion), their two nearest neighbors will adopt their opinion (ferromagnetic-like). However, if the pairs of agents in the row and column directions do not agree, their two nearest neighbors will adopt alternating opinions (antiferromagnetic-like) \cite{sznajd2000opinion}, as shown in Fig.~\ref{fig:2D_square} (b).
        \end{enumerate}
    \end{enumerate}
\end{itemize}

The difference between Case 1 and Case 2 is that in Case 1, the eight neighbors adopt the opinion of the four influencer agents only if all four agents share the same opinion. 
In Case 2, if the four influencer agents do not share the same opinion, the eight nearest neighboring agents can still change their opinions. If the nearest pair of influencer agents in the row and column directions have the same opinion, then their two nearest neighbors will adopt this opinion (ferromagnetic-like). Conversely, these two nearest neighbors will adopt alternating opinions if the pairs of agents in the row and column directions do not share the same opinion (antiferromagnetic-like). For example, if only two agents share the same opinion, such as \( S_{(i,j)} = S_{(i+1,j)} \), then their two neighbors will adopt this opinion, resulting in \( S_{(i-1,j)} = S_{(i,j)} = S_{(i+1,j)} = S_{(i+2,j)} \). Conversely, if two agents do not share the same opinion, such as \( S_{(i,j-1)} \neq S_{(i+1,j-1)} \), then their two neighbors will adopt alternating opinions, resulting in \( S_{(i,j-1)} = S_{(i+2,j-1)} \) and \( S_{(i-1,j-1)} = S_{(i+1,j-1)} \).
This scheme is consistent with the one-dimensional Sznajd model \cite{sznajd2000opinion}. For the model on a 2D lattice, we apply periodic boundary conditions.

The order parameter (magnetization) of the system can be computed analytically using 
\begin{equation}\label{eq:order_th}
m = \dfrac{1}{N} \sum \sigma_i.
\end{equation}
In the Monte Carlo simulation, we use $\langle m \rangle = 1/R \sum_{i =1}^{R} m_i$, where $\langle \cdots \rangle$ is the average of all samples. We also estimate the critical exponents of the model to define the universality class using the finite-size scaling relation as follows \cite{cardy1996scaling}:
\begin{align}
	m ( N) & \sim N^{-\beta/\nu}, \label{eq1} \\
	\chi(N) & \sim N^{\gamma/\nu}, \label{eq2}\\
	U(N) &\sim \text{constant}, \label{eq3}\\
	p_{c}(N)-p_{c} & \sim N^{-1/\nu}, \label{eq4}
\end{align}
where $\chi$ and $U$ are the susceptibility and Binder cumulant, respectively, defined as:
\begin{align}
	\chi =& N \left( \langle m^2 \rangle - \langle m \rangle^2\right), \label{Eq:susc}\\
	U =& 1 - \dfrac{\langle m^4 \rangle}{3\, \langle m^2 \rangle^2}. \label{Eq:Binder}
\end{align}
The scaling parameters work near the critical point of the system.

\section{Results and discussion}
\label{sec.3}
\subsection{Model on the complete graph}
In the complete graph, all agents are connected with the same probability, meaning all nodes and links are homogeneous and isotropic. This concept is analogous to the mean-field theory in statistical physics~\cite{amit2005field}, which implies that all fluctuations in the system can be ignored. Consequently, we can consider all agents as neighbors, each having \( N-1 \) neighbors. To describe the system's state, we define the fraction of agents with the opinion `up' as \( c = N_{\uparrow}/N \), where \( N \) is the total population, \( N_{\uparrow} \) is the number of agents with the opinion `up,' and \( N_{\downarrow} \) is the number of agents with the opinion `down.' Therefore, \( N = N_{\uparrow} + N_{\downarrow} \). During the dynamics process, the fraction \( c \) increases with probability \( \rho^{+} \) and decreases with probability \( \rho^{-} \), while remaining constant with probability \( \left(1 - \rho^{+} - \rho^{-}\right) \). The explicit forms of \( \rho^{+} \) and \( \rho^{-} \) depend on the specific model being considered.

As stated in the original Sznajd model~\cite{sznajd2000opinion}, two paired agents in a one-dimensional lattice with the same opinion will influence their neighbors to adopt their opinion, a process known as social validation. The final state of this interaction is homogeneous, akin to ferromagnetism in statistical physics. If the two paired agents hold different opinions, their neighbors will adopt the opposite opinion, resulting in a completely disordered state, reflecting an antiferromagnetic character. Extending the original Sznajd model, we can say that groups of three, four, or more paired agents with the same opinion will influence a group of their nearest neighbors to adopt their opinion. If a paired agent has only one neighbor to influence, the Sznajd model becomes equivalent to the nonlinear \( q \)-voter model~\cite{castellano2009nonlinear}. Therefore, in the complete graph, we consider several paired agents chosen randomly, each influencing one, two, or three of their neighbors, also chosen randomly, following the interaction rules based on the algorithm mentioned previously.

In a single time step, the maximum change in the fraction of opinions \( c \) will be \( \pm n/N \). Thus, we can formulate the general expression for the probability of the opinion \( c \) increasing and decreasing for any group of paired agents \( r \) and their neighbors \( n \), which can be written as:
\begin{equation}\label{eq:prob}
    \begin{aligned}
        \rho^{+}(N,r,p,n) =&  N_{\downarrow}\left[ \dfrac{n\left(1-p\right) \prod_{j =1}^{r} \left(N_{\uparrow}-j+1\right)}{\prod_{j=1}^{r+1}\left(N-j+1\right)} +\dfrac{np}{2\,N} \right], \\
        \rho^{-}(N,r,p,n) =& N_{\uparrow}\left[\dfrac{n\left(1-p\right) \prod_{j =1}^{r} \left(N_{\downarrow}-j+1\right)}{\prod_{j=1}^{r+1}\left(N-j+1\right)} +\dfrac{np}{2\,N} \right].
    \end{aligned}
\end{equation}
For \( r = 2 \) and \( n = 2 \), the model is reduced to the original Sznajd model on the complete graph. For \( r \geq 2 \) and \( n = 1 \), the model is also reduced to the nonlinear \( q \)-voter model with independence as discussed in Ref.~\cite{nyczka2012phase}, and the nonlinear \( q \)-voter model with independence with \( s = 1 \) as discussed in Ref.~\cite{muslim2023effect}. Since the model is defined on the complete graph, it is suitable for a large system size \( N \gg 1 \). Consequently, Eq.~\eqref{eq:prob} simplifies to the following forms:
\begin{equation}\label{eq:prob1}
    \begin{aligned}
        \rho^{+}(c,r,p,n) = & \left(1-c\right)\left[n\left(1-p\right)c^r + \frac{np}{2} \right], \\
        \rho^{-}(c,r,p,n) = & c\left[n\left(1-p\right)\left(1-c\right)^r+ \frac{np}{2}\right],
    \end{aligned}
\end{equation}
Eq.~\eqref{eq:prob1} is the essential equation for analyzing various macroscopic phenomena of systems, such as the occurrence of the order-disorder phase transition in this model.

\subsection{Time evolution and steady state}
The temporal evolution of the fraction opinion \( c \) can be analyzed using the following recursive formula \cite{krapivsky2010kinetic}:
\begin{equation}\label{eq:time_evol_c}
    c(t') = c(t) + \dfrac{1}{N} \left(\rho^{+}(c,r,p,n) -\rho^{-}(c,r,p,n)\right),
\end{equation}
where the time evolution is measured in sampling events corresponding to the Monte Carlo steps. To align Eq.~\eqref{eq:time_evol_c} with the Monte Carlo simulation, the fraction opinion \( c \) needs to be measured in a Monte Carlo sweep by rescaling \( t \) by a factor of \( 1/N \). In other words, one Monte Carlo step is \(\delta t = 1/N\), as a single Monte Carlo sweep corresponds to \(\delta t \, N = 1\). In the limit of a large population size, where \(N \to \infty\) or \(\delta t \to 0\), Eq.~\eqref{eq:time_evol_c} can be expressed in differential form as:
\begin{equation}\label{eq:time_ev_diff}
\dfrac{\mathrm{d}c}{\mathrm{d}t} =  \rho^{+}(c,r,p,n) -\rho^{-}(c,r,p,n). 
\end{equation}

Theoretically, we can obtain the exact solution for the fraction opinion \( c \) over time \( t \) by substituting Eq.~\eqref{eq:prob1} into Eq.~\eqref{eq:time_ev_diff} and integrating it. However, finding the exact solution for the fraction opinion \( c \) is challenging for any paired agent size \( r \). For simplicity, in the case of the original Sznajd model with \( r = 2 \), the solution of Eq.~\eqref{eq:time_ev_diff} can be written as:
\begin{equation}\label{eq:c_exact}
    c(t,p,n)= \dfrac{1}{2}  + \dfrac{1}{2}\left( \dfrac{1-3p}{1-p+2\exp\left[-n\left(1-3p\right)\left(t + A\right)\right]}\right)^{1/2},
\end{equation}
where $A$ is a parameter that satisfies the initial condition $t = 0, \, c (t) =  c_0$, namely, $A = \ln[(2\,c_0-1)^2/2\,(1-p)\,(c_0-c_0^2)-p]/n(1-3p)$. Based on Eq.~\eqref{eq:c_exact}, we can observe that \( c(t,n,p) \) will evolve to two steady states \( c_{1,2} \) for \( p < 1/3 \) and one stationary state \( c = 1/2 \) for \( p \to 1/3 \) for any values of neighbor size \( n \) and initial fraction opinion \( c_0 \). In this case, \( p = 1/3 \) is the critical point that causes the model to undergo an order-disorder phase transition. In the next section, we will determine the model's critical point for any values of \( r \) by considering the stationary condition of Eq.~\eqref{eq:time_ev_diff}.

\begin{figure}[tb]
    \centering
    \includegraphics[width = \linewidth]{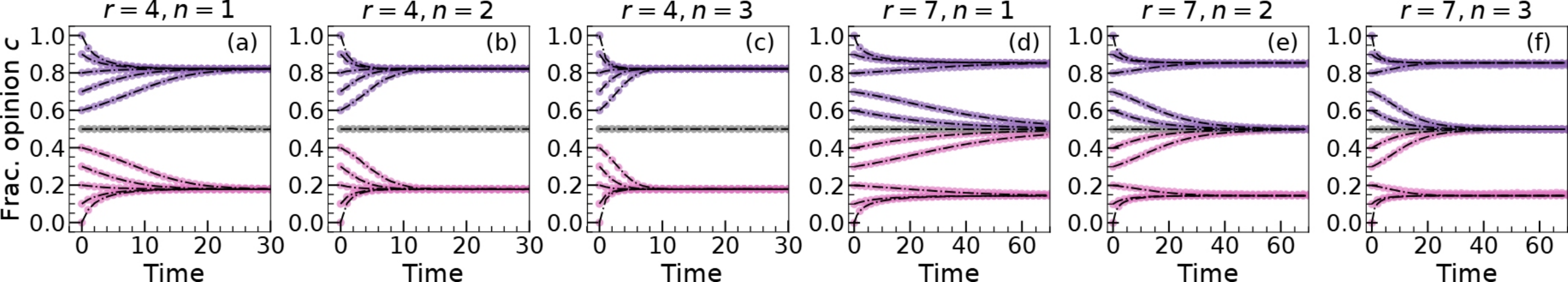}
    \caption{Time evolution of the fraction opinion $c$ of the model with four paired agents with one to three of their neighbors [panels (a), (b), and (c)] for the same $p = 0.2$. Panels (d), (e), and (f) are for seven paired agents with one to three of their neighbors for the same $p = 0.12$. As seen, the fraction opinion $c$ evolves to the same $c_{st}$ for the same $r$, namely two stables $c_{st}$ for $r = 4$, three stables $c_{st}$ for $r = 7$. Data points and dashed lines represent numerical simulation and Eq.~\eqref{eq:time_evol_c}, respectively. Population size $N = 10^5$, and each data point averages over $500$ independent realizations. }
    \label{fig:time_evol_one}
\end{figure}

We are particularly interested in solving Eq.~\eqref{eq:time_ev_diff} for any values of paired agent size \( r \) using numerical methods, such as the Runge-Kutta 4th order method~\cite{pinder2018numerical}, and comparing these results with Monte Carlo simulations. For instance, for \( r = 4, 7 \) and \( n = 1, 2, 3 \), the time evolution of the fraction opinion \( c(t) \) is shown in Fig.~\ref{fig:time_evol_one}. From the figure, it is evident that for the same \( r \), the fraction opinion \( c \) evolves to the same stable or stationary value \( c_{\text{st}} \) for all \( n = 1, 2, 3 \). This result indicates that the stationary fraction \( c_{\text{st}} \) is not affected by the neighbor size \( n \). This result also indicates that the critical point depends only on the number of paired agents, size \( r \). However, there is a difference in the time required for the fraction opinion \( c(t) \) to reach a steady state; specifically, the time needed to reach a steady state is inversely proportional to the number of neighbor sizes \( n \), i.e., \( t_{\text{steady}} \sim 1/n \) for example, for \( r = 2 \), we can determine the explicit form of \( t_{\text{steady}} \) by substituting the value of \( A \) into Eq.~\eqref{eq:c_exact} and considering \( c_{t_{\text{steady}}} = \text{constant} \)). It is easy to understand that when the number of interacting agents increases at a time step $t$, the time needed to reach the entire interacting population is faster, affecting the time to reach a steady state faster.

\begin{figure}[tb]
    \centering
    \includegraphics[width = 0.5\linewidth]{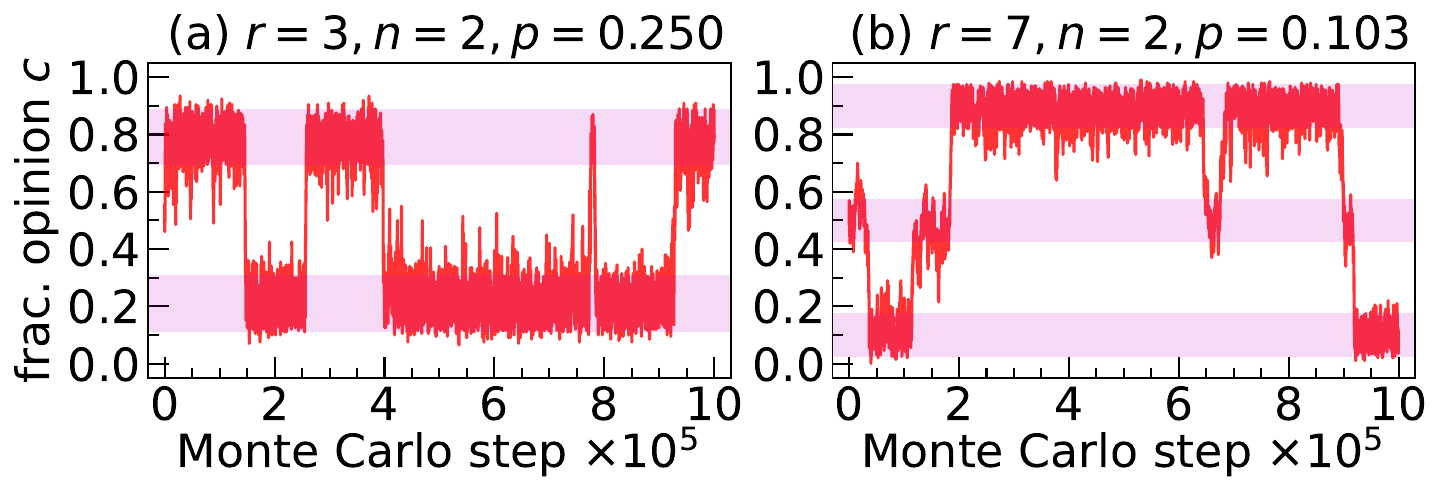}
    \caption{Time evolution of the fraction opinion $c$ per site for the model with $r = 3, n = 2, p = 0.25$ [panel (a)], and  $r = 7, n = 2, p = 0.103$ [panel (b)]. As seen in panel (a), the fraction opinion $c$ fluctuates at two stables states at $c_{1,2} \neq 0.5$ (colored regions), indicating the model undergoes a second-order phase transition, while in panel (b), the fraction opinion $c$ fluctuates at three stable states at $c_{1,2} \neq 0.5$ and $c_{3} = 0.5$ (colored regions), indicating the model undergoes a first-order phase transition.}
    \label{fig:mcs}
\end{figure}

As mentioned previously, we can also analyze the existence of the order-disorder phase transition of the model through the fluctuation behavior of the fraction opinion \( c \) at each time step (Monte Carlo step), as shown in Fig.~\ref{fig:mcs} for cases \( r = 3, 7 \) and the same \( n = 2 \). We observe that the fraction opinion \( c \) fluctuates between two bistable states \( c_{1,2} \neq 0.5 \) for the model with \( r = 4 \) and \( p < p_c = 1/3 \), and three stable states \( c_{1,2} \neq 0.5 \) and \( c_{3} = 0.5 \) for \( p > p_c = 3/35 \). The two stable states correspond to the occurrence of a second-order (continuous) phase transition, while the three stable states correspond to a first-order (discontinuous) phase transition. We find the same phenomenon for all \( r \leq 5, n \neq 0 \) (two stable states) and all \( r > 5, n \neq 0 \) (three stable states).

We will see more clearly by considering the equilibrium condition of Eq.~\eqref{eq:time_evol_c} as follows:

\begin{equation}\label{eq:critical}
    p = \dfrac{c_{st}\left(1 - c_{st}\right)^r + c_{st}^{1 + r} - c_{st}^r}{c_{st}\left(1 - c_{st}\right)^r + c_{st}^{1 + r} - c_{st}^r - c_{st} + 1/2},
\end{equation}
where  \( p_c \) is the critical point that causes the model to undergo an order-disorder phase transition is obtained by setting the limit \( c \to 1/2 \), resulting in \( p_c = p_{\lim c \to 1/2} \). Eq.~\eqref{eq:critical} is essentially the same as the \( q \)-voter model with independence \cite{nyczka2012phase}. As previously mentioned, the critical point \( p_c \) or the stationary value \( c_{st} \) is not influenced by the number of neighbors \( n \) but only by the paired agents' size \( r \), as shown in Eq.~\eqref{eq:critical}. In terms of the order parameter \( m \), Eq.~\eqref{eq:critical} can be written as \( m \sim (p - p_c)^{\beta} = (p - p_c)^{1/2} \), where \( m = 2c_{st} - 1 \) and \( \beta = 1/2 \) is the critical exponent for \( r \leq 5 \), causing all data \( N \) to collapse near the critical point \( p_c \). In the next section, we will determine other critical exponents, \( \nu \) and \( \gamma \), corresponding to the Binder cumulant \( U \) and susceptibility \( \chi \), using Monte Carlo simulations.

\begin{figure}[tb]
    \centering
    \includegraphics[width = 0.35\linewidth]{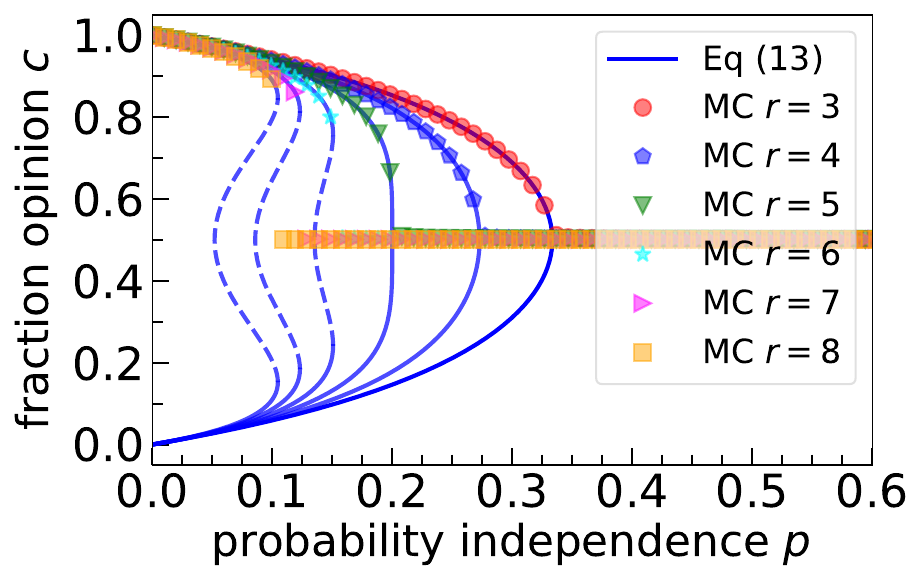}
    \caption{(Phase diagram) The comparison between Eq.~\eqref{eq:critical} (solid lines) versus Monte Carlo simulation (data points) for several values of paired agent size $r$ and showing the well-agreement result. As seen, the model undergoes a second-order (continuous) phase transition for $r \leq 5$ and a first-order (discontinuous) phase transition for $r > 5$. Dashed lines represent the imaginary part of $c_{st}$. The population size $N = 10^5$, and each data point averages $10^6$ independent realizations.}
    \label{fig:phase_diagram}
\end{figure}

Fig.~\ref{fig:phase_diagram} shows the comparison between Eq.~\eqref{eq:critical} and the Monte Carlo simulation, demonstrating a solid agreement. The model undergoes a second-order (continuous) phase transition for \( r \leq 5 \) and a first-order (discontinuous) phase transition for \( r > 5 \) (solid-dashed lines). The dashed lines represent the imaginary part of Eq.~\eqref{eq:critical}. This data can be correlated with the time evolution of the fraction opinion \( c \), as illustrated in Fig.~\ref{fig:time_evol_one}. For instance, when \( r = 4 \), the fraction opinion \( c \) evolves to two stable states (continuous state), while for \( r = 7 \), the evolution of the fraction opinion \( c \) evolves to three stable states (discontinuous state). Furthermore, we analyze the occurrences of continuous and discontinuous phase transitions within the model by examining the stationary probability density function of \( c \) and the effective potential, as elaborated in the subsequent section.

\subsection{Effective potential and Landau paradigm}
The order-disorder phase transition of a model can also be analyzed using the system's effective potential-like, which is defined as:
\begin{equation}\label{eq:pot}
    V_{\text{eff}} = -\int F_{\text{eff}} \, \mathrm{d}c
\end{equation}
where \( F_{\text{eff}} = (\rho^{+} - \rho^{-}) \) is the effective force that flips the spin during the dynamics process. The effective potential in Eq.~\eqref{eq:pot} is used to analyze the movement of public opinion in a bistable potential~\cite{nyczka2012opinion}. In this paper, we examine the dynamics of public opinion within the framework of both a two-stable-state potential (bistable potential) and a three-stable-state potential (three-stable potential), utilizing the model's characteristics. By substituting Eq.~\eqref{eq:prob1} into Eq.~\eqref{eq:pot} and subsequently integrating, the effective potential \( V_{\text{eff}} \) for the model is obtained as follows:
\begin{align}\label{eq:pot_eff}
     V_{\text{eff}}(n,c,r,p) =& n\left(1-p\right)\Big[\dfrac{c^{r+2}}{r+2}-\dfrac{c^{r+1}}{r+1}-\dfrac{\left(c\,r+c+1\right)}{\left(r+1\right)\left(r+2\right)} \left(1-c\right)^{r+1} \Big] -c\left(1-c\right)\dfrac{np}{2}.
\end{align}
One can check that for all values \( n > 0 \) and \( r > 1 \), when \( p = 0 \) (there are no independent agents), the potential is bistable at \( c_{1,2} = 0, 1 \) and unstable at \( c_3 = 1/2 \). This condition implies that all opinions are in a complete consensus (completely ordered state), where all agents share the same opinion. In contrast, for \( p = 1 \), the effective potential is in a monostable state at \( c = 1/2 \). This condition indicates that all agents are in a completely disordered state. Note that the potential in Eq.~\eqref{eq:pot_eff} depends on the parameter \( n \), and for \( n = 1 \), Eq.~\eqref{eq:pot_eff} becomes identical to the potential of the \( q \)-voter model with independence as mentioned in Ref.~\cite{nyczka2012phase}.

\begin{figure}[tb]
    \centering
    \includegraphics[width = 0.5\linewidth]{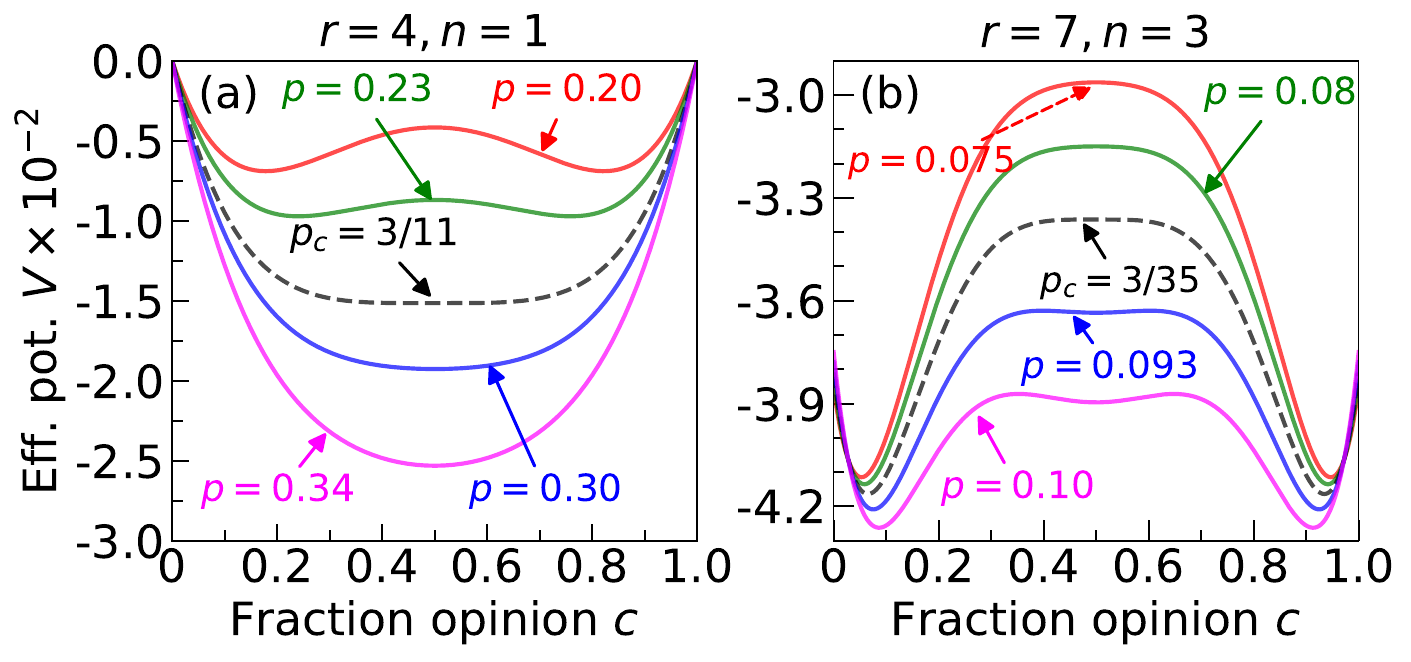}
    \caption{The effective potential $V(n,c,r,p)$ of the model is based on Eq.~\eqref{eq:pot_eff} for $r = 4, n = 1$ [panel (a)] and $r = 7, n = 3$ for $p < p_c, p = p_c$ and $p > p_c$. As seen for both panels, there are bistable states for $ p < p_c$ at $c_{1,2} = c_{st}$, and unstable at $c = 1/2$. For the case $r = 4, n =1$, only one monostable state for $p > p_c$ at $c= 1/2$, and for the case $r  = 7, n =3$, there are three stable states for $p > p_c$ near the critical point. For both panels, the transition bistable-monostable (three stable) at $p=p_c$ (dashed line) indicates the model undergoes a second-order phase transition for the case $r = 4, n = 1$ and a first-order phase transition for the case $r = 7, n = 3$.}
    \label{fig:pot}
\end{figure}

To visualize Eq.~\eqref{eq:pot_eff}, we plot it for typical values of \( r \) and \( n \) as shown in Fig.~\ref{fig:pot}. In panel (a), for \( r = 4 \) and \( n = 1 \), the effective potential \( V_{\text{eff}} \) is bistable for \( p < p_c \) and monostable for \( p > p_c \), indicating that the model undergoes a second-order phase transition. From a social perspective, this means that when the independent behavior \( p \) in the population is low, all agents are less likely to change their opinion from up to down or vice versa. In other words, in this situation, all agents will tend to maintain their opinions. As \( p \) increases, the likelihood of agents changing their opinions increases, leading to a status quo or stalemate situation at the critical independence \( p_c \). In panel (b), for \( r = 7 \) and \( n = 3 \), the effective potential exhibits a different character than in panel (a). Here, the potential has bistable states for \( p < p_c \) and three stable states for \( p > p_c \) near the critical point \( p_c \), indicating that the model undergoes a first-order phase transition. This scenario reflects a more complex social dynamic where higher levels of independent behavior lead to multiple stable states in the system.

The stationary $c_{\text{st}}$ that makes the effective potential in Eq.~\eqref{eq:pot_eff} to be maximum and minimum is given by Eq.~\eqref{eq:critical}. The critical point that makes the model undergo an order-disorder phase transition can be analyzed from the transition maximum-minimum of the effective potential $V_{\text{eff}}$, that is, by 
$\mathrm{d}^2V_{\text{eff}}/\mathrm{d}c^2|_{c = 1/2} = 0$:
\begin{equation}\label{eq:critical_p}
    p_c(r) = \dfrac{r^2+r-2}{r^2+r-2+2^r+r2^{r-1}}.
\end{equation}
Eq.~\eqref{eq:critical_p} is the same as Eq.~\eqref{eq:critical} for limit $c \to 1/2$.

The order-disorder phase transition of the model can also be analyzed using the Landau potential. In classical Landau theory concerning phase transitions~\cite{landau1937theory, plischke1994equilibrium}, Landau posited that the free energy could be expanded in a power series near the critical point, expressed in terms of the order parameter. The Landau potential applies not only to equilibrium systems described by thermodynamic parameters such as pressure, temperature, volume, and other properties. However, it can also be extended to analyze nonequilibrium systems, as exemplified in the Langevin equation for two absorbing states using a mean-field approximation \cite{al2005langevin,vazquez2008systems}. In general, the Landau potential may depend on the system's thermodynamic parameters and order parameters, as encapsulated in Eq.~\eqref{eq:order_th}. In this context, we employ Landau's theory to scrutinize the order-disorder phase transition of the model. Consequently, the potential \(V\) is expressed as:
\begin{equation} \label{eq:pot_land}
    V = \sum_i V_i m^i =  V_0 + V_1 m + V_2 m^2  + V_3 m^3 + V_4 m^4 + \cdots.
\end{equation}
Note that the potential \( V \) in Eq.~\eqref{eq:pot_land} is symmetric under inversion \( m \to -m \); therefore, the odd terms vanish. The term \( V_i \) can depend on the probability of independence \( p \) and the paired agents' size \( r \), which are the essential parameters in this model. 

\begin{figure}[tb]
    \centering
    \includegraphics[width = 0.35\linewidth]{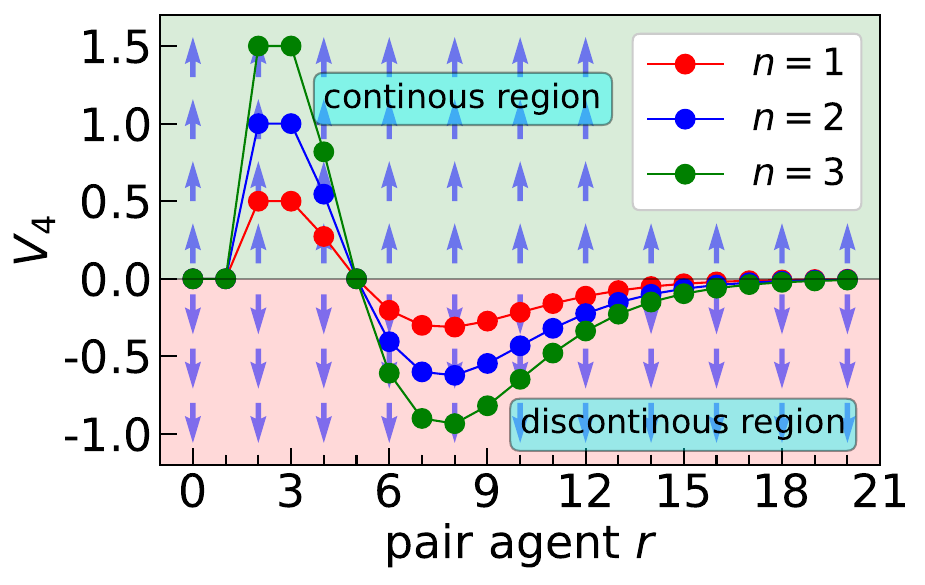}
    \caption{The plot of Eq.~\eqref{eq:parV4} for several values $n$. As seen, the parameter $V_4 \geq 0 $ for $ r \leq 5$ and $V < 0$ for $r > 5$ for all values $n$, indicate the model undergoes a second-order (continuous) phase transition for $ r \leq 5$ and a first-order (discontinuous) phase transition for $ r > 5$ for all values $n$.}
    \label{fig:v4}
\end{figure}

We can simplify the analysis of phase transitions in the model by considering only two terms, namely \( V = V_2 m^2 + V_4 m^4 \). Based on Eq.~\eqref{eq:pot_land}, for \( V_2 < 0 \), the potential is in a bistable state, and for \( V_2 > 0 \), the potential is in a monostable state. The phase transition can be characterized by the condition \( V_2 = 0 \). Thus, by comparing Eqs.~\eqref{eq:pot_eff} (after rescaling \( c = (m+1)/2 \)) and \eqref{eq:pot_land}, we obtain \( V_2 \) and \( V_4 \) for the model as follows:
\begin{equation}
\begin{aligned}
    V_2 (n,r,p) = \dfrac{pn}{2}-\dfrac{n\left(1-p\right)}{r+2} \left(r^2+r-2\right),
\end{aligned}
\end{equation}
and by setting $V_2 = 0$, the critical point $p_c$ of the model is given by:
\begin{equation}\label{eq:critical_point}
    p_c(r) = \dfrac{r^2+r-2}{r^2+r-2+2^r+r2^{r-1}}.
\end{equation}
We have the same formula as Eq.~\eqref{eq:critical_p}. We also obtain $V_4$ as:
\begin{equation}\label{eq:parV4}
    V_4(n,p) = \dfrac{-nr\left(r - 1\right)\left(r + 2\right) \left(r - 5\right)}{2^{r + 1} \left(r + 2\right)+ 4\left(r^2 + r - 2\right)}.
\end{equation}
Eq.~\eqref{eq:parV4} is very important to recognize the boundary, whether a continuous or discontinuous phase transition occurs. The plot of Eq.~\eqref{eq:parV4} as shown in Fig.~\ref{fig:v4}. It can be seen that $V_4$ is positive for $ r \leq 5$ for all values $n$, indicating the occurrence of the second-order phase transition (continuous region), while $V_4$ is negative for $r > 5$ for all values $n$, indicating the occurrence of the first-order phase transition (discontinuous region).

\subsection{Probability density function}
\begin{figure}[tb]
    \centering
    \includegraphics[width = 0.6\linewidth]{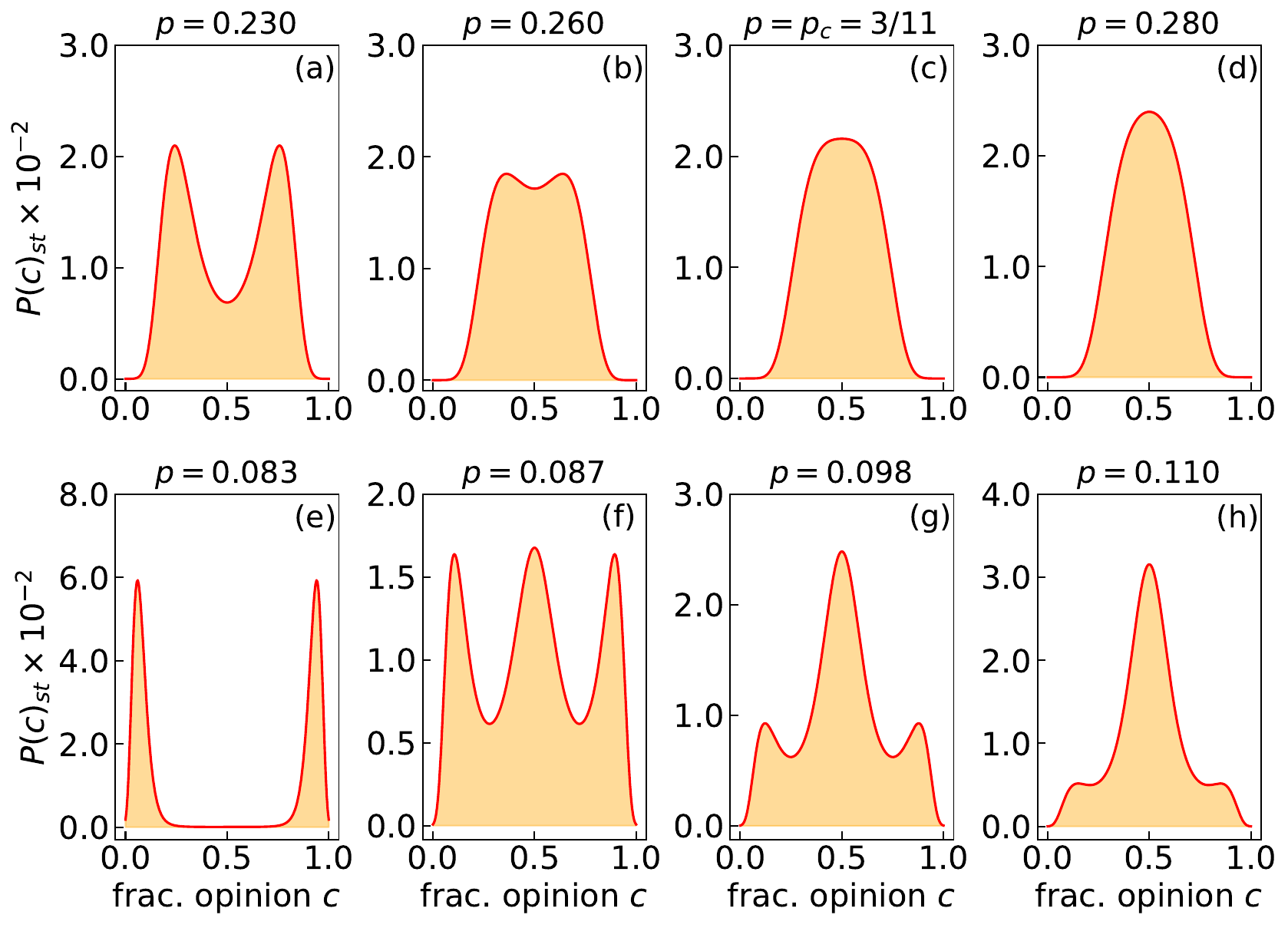}
    \caption{The probability density function of fraction opinion $c$ in Eq.~\eqref{fokker-planck_solution} for several values of independence $p$. Panels (a) - (d) for the model with $r = 4$ and panels (e) - (h) for the model with $r = 7$. It can be seen for both $r$ that for $p < p_c$, there are two peaks of $P(c)_{st}$ indicating the system in two stable states. For $p > p_c$, the probability density $P(c)_{st}$ only has one peak at $c = 1/2$ for the model with $r = 4$ and three peaks for the model with $r = 7$, indicating the model with $r = 4$ and $r = 7$ undergoes a continuous and discontinuous phase transition, respectively.}
    \label{fig:one_ne}
\end{figure}

We can analyze the order-disorder phase transition of the model through the stationary probability density function of the spin-up fraction \( c \). Generally, the differential equation for the probability density function \( P(c,t) \) of the fraction \( c \) at time \( t \) can be approximated using the Fokker-Planck equation as follows \cite{frank2005nonlinear}:
\begin{equation}\label{eq.fokker-planck}
    \dfrac{\partial P(c,t)}{\partial t} =-\dfrac{\partial}{\partial c} \left[\xi_1(c)P(c,t) \right]+\dfrac{1}{2} \dfrac{\partial^2}{\partial c^2} \left[ \xi_2(c) P(c,t)\right].
\end{equation}
Thus, the general solution for the stationary condition of Eq.~\eqref{eq.fokker-planck} can be written as:
\begin{equation}\label{fokker-planck_solution}
 P(c)_{st} = \dfrac{C}{\xi_2} \exp\left[\int 2\dfrac{\xi_1}{\xi_2}\mathrm{d}c \right],   
\end{equation}
where $C$ is the normalization constant that satisfies $\int_{0}^{1} P(c)_{st} \, \mathrm{d}c = 1$. The parameters $\xi_1$ and $\xi_2$ can be considered as the drift-like  and diffusion-like coefficients, which are defined as:
\begin{equation}
\begin{aligned}
    \xi_1 & = \frac{1}{2}\left[\rho^{+}(c,r,p,n)+\rho^{-}(c,r,p,n)\right] \\
    \xi_2 & = \left[\rho^{+}(c,r,p,n)-\rho^{-}(c,r,p,n)\right], 
\end{aligned}    
\end{equation}
or explicitly can be written as:
\begin{equation} \label{eq:drift-diffusion}
\begin{aligned}
    \xi_1 & = \frac{\left(1-c\right)n}{2} \left[ c^r \left(1-p\right) +  \dfrac{p}{2} \right] + \frac{cn}{2} \left[ \left(1-c\right)^r \left(1-p\right) +  \dfrac{p}{2} \right], \\
    \xi_2 & = {\left(1-c\right)n}\left[ c^r \left(1-p\right) +  \dfrac{p}{2} \right] - {cn} \left[ \left(1-c\right)^r \left(1-p\right) +  \dfrac{p}{2} \right].
\end{aligned}
\end{equation}
One can see that obtaining the exact solution of Eq.~\eqref{fokker-planck_solution} can take much work. We are interested in solving it numerically. The plot of Eq.~\eqref{eq.fokker-planck} for the model with $r = 4$ and $r = 7$ is exhibited in Fig.~\ref{fig:one_ne}. Similar to the effective potential for the model with $r = 4$ and $ r = 7$, there are two peaks at $c_{1,2} = c_{st}$, while for $p > p_c$ there is one peak at $c = 1/2$ for $r=4$ and three peaks for $r = 7$. The behavior of $P_{st}$ in the model for $r = 4$ and $r = 7$ is typical of a system undergoing continuous and discontinuous phase transitions, respectively.

\subsection{Critical exponents and universality class}
\textbf{\textit{The model on the complete graph}}. This section delves into the analysis of critical points and exponents that yield the optimal collapse of all data, focusing on the model exhibiting a second-order phase transition, specifically for \( r \leq 5 \) and for any \( n \neq 0 \). By employing the finite-size scaling relations in Eqs.~\eqref{eq1} - \eqref{eq4}, we deduce the critical exponents of the model are \( \beta \approx 0.5 \), \( \gamma \approx 1.0 \), and \( \nu \approx 2.0 \) (not displayed), with the critical point determined by Eq.~\eqref{eq:critical_p}. These critical exponents demonstrate universality, consistently yielding the same values for all \( N \). Notably, the critical exponents \( \beta = 1/2 \) and \( \gamma = 1 \) are the same as typical mean-field exponents, deviating only in the case of \( \nu = 2 \). However, this discrepancy is attributed to a higher critical dimension of \( d_c = 4 \), yielding an effective exponent \( \nu' = 1/2 \), such that \( \nu = d_c \nu' = 2 \). Based on the data, our findings suggest that the model falls into the same universality class as the \( q \)-voter model~\cite{muslim2023effect}, the kinetic exchange model~\cite{crokidakis2014phase}, various complex network models~\cite{hong2007finite,mulya2024phase}, and remains within the mean-field Ising universality class.

\begin{figure}[tb]
    \centering
    \includegraphics[width = 0.85\linewidth]{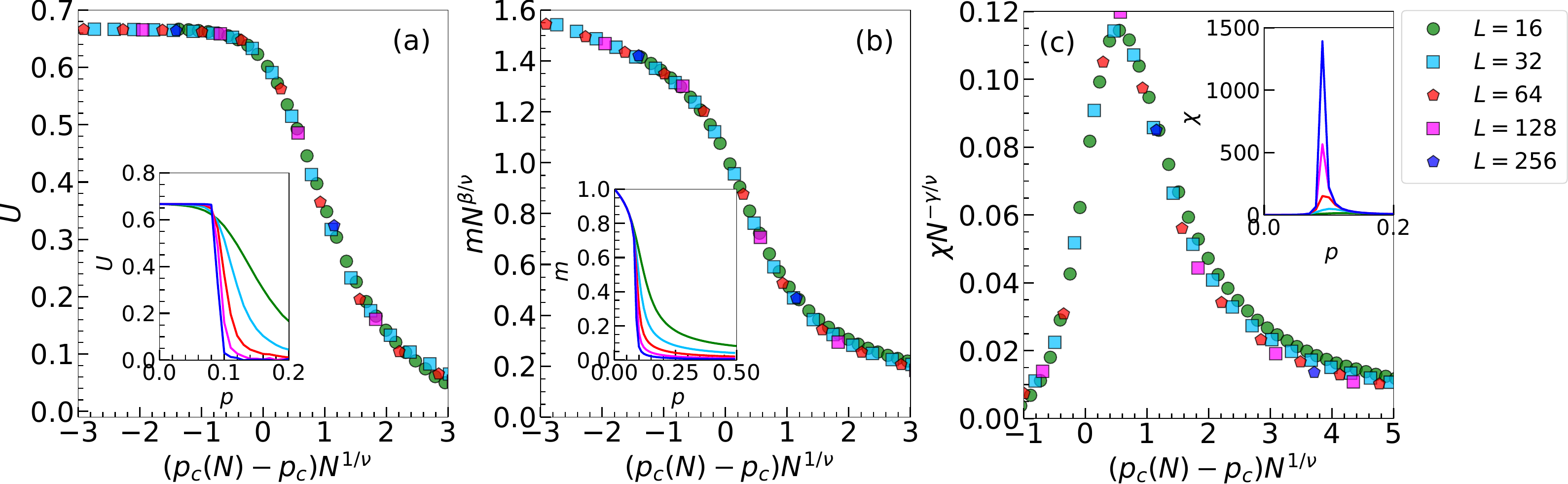}
    \caption{Continuous phase transition of the Sznajd model on the two-dimensional square lattice with only four homogeneous agents (case 1). The critical point is obtained from the cross of lines between Binder cumulant $U$ versus probability independence $p$ that occurred at $p_c \approx 0.0855$ [inset graph (a)]. The critical exponents that make the best collapse of all data are $\gamma \approx 1.75, \beta \approx 0.125,$ and $\nu \approx 1.0$ (main graph).}
    \label{fig:2d-four-only}
\end{figure}

\textbf{\textit{The model on the two-dimensional square lattice}}. On the square lattice, we explore various values of the linear lattice size \( L \), specifically \( L = 16, 32, 64, 128, 256 \), and calculate the parameters \( m \), \( \chi \), and \( U \), as defined in Eqs.~\eqref{Eq:susc} - \eqref{Eq:Binder}. Each data point is obtained from an average over \( 3 \times 10^6 \) independent realizations to obtain the best results. The Monte Carlo simulation outcome for the model in case (1), involving only four paired agents with the same opinion, is presented in Fig.~\ref{fig:2d-four-only}. One can see that the model undergoes a continuous phase transition, with the critical point determined by the intersection of lines in the Binder cumulant \( U \) versus the probability of independence \( p \), occurring at \( p_c \approx 0.0805 \) [inset graph (a)]. The main graphs show the scaling plot of the model using finite-size scaling analysis, leading to critical exponents that achieve the optimal collapse of all data, specifically \( \beta \approx 0.125 \), \( \gamma \approx 1.75 \), and \( \nu \approx 1.0 \). Based on these scaling values, the universality class of the model falls to the two-dimensional Ising model universality class.

\begin{figure}[tb]
    \centering
    \includegraphics[width = 0.65\linewidth]{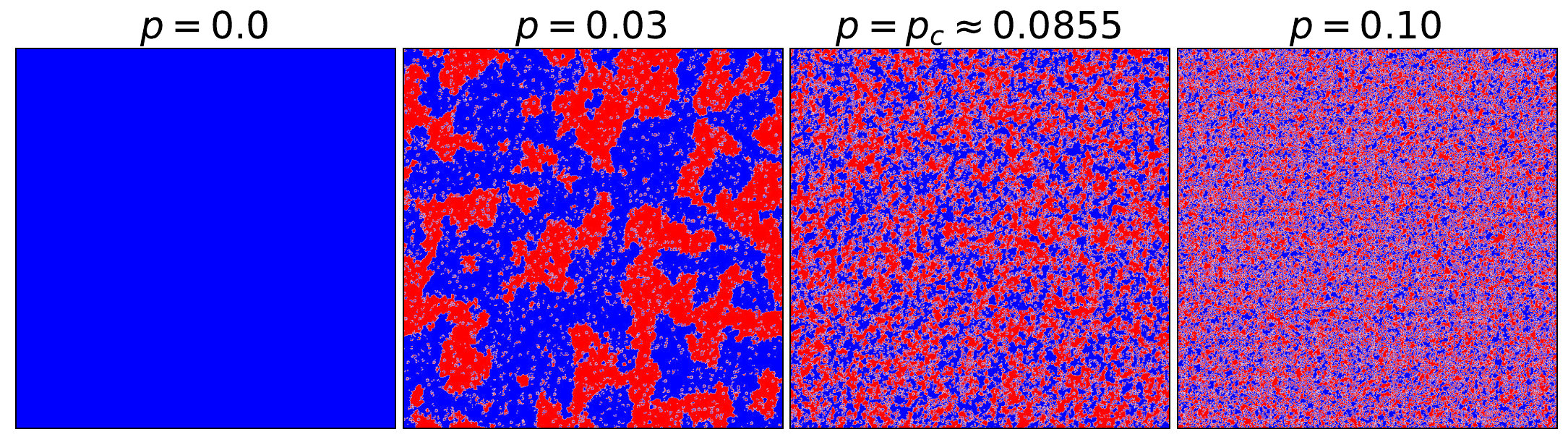}
     \caption{Snapshot of the dynamics of agents' interaction in an equilibrium state of the model with independent agents on the two-dimensional square lattice for typical probability independence $p$. From left to right  $p = 0.0, p = 0.03 , p = p_c$, and $p = 0.10$. The linear square lattice size $L = 512$.}
    \label{fig9:snapshot}
\end{figure}

Fig.~\ref{fig9:snapshot} shows snapshots of the model at equilibrium state for typical values of the probability of independence \( p \). The initial state is disordered, with an equal number of opinions up and down. From left to right, the panels represent \( p = 0.0 \), below the critical point, at the critical point \( p_c \approx 0.0805 \), and above the critical point. As shown, at \( p = 0.0 \) (no independent agents), the system is in a homogeneous state (complete consensus), meaning all agents share the same opinion, either up or down, with the absolute value of the order parameter \( m = 1.0 \). At the critical point \( p_c \), the system approaches a completely disordered state, with magnetization close to zero. Above the critical point, the system becomes completely disordered (stalemate situation).

\begin{figure}[tb]
    \centering
    \includegraphics[width = 0.85\linewidth]{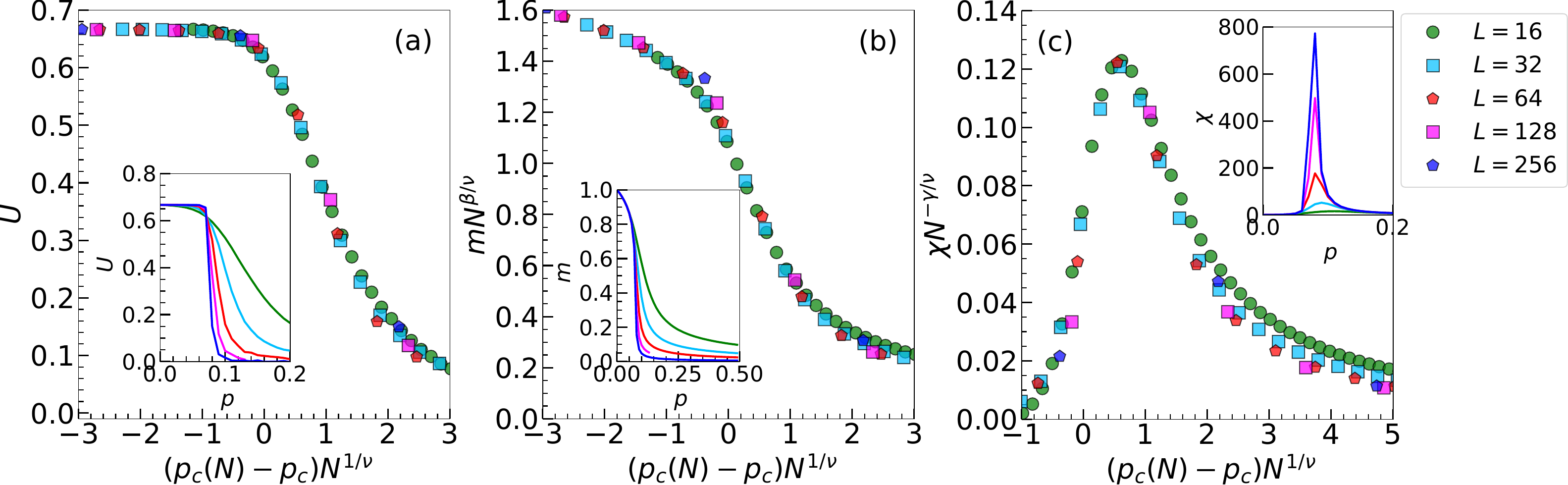}
    \caption{Continuous phase transition of the Sznajd model on the two-dimensional square lattice with not only four homogeneous agents (case 2). The critical point is obtained from the cross of lines between Binder cumulant $U$ versus probability independence $p$ that occurred at $p_c \approx 0.0715$ [inset graph (a)]. The critical exponents that make the best collapse of all data are $\gamma \approx 1.75, \beta \approx 0.125,$ and $\nu \approx 1.0$ (main graph).}
    \label{fig:2d-not-homogen}
\end{figure}

For the model in case (2), where not only four paired agents share the same opinion, the Monte Carlo simulation results are shown in Fig.~\ref{fig:2d-not-homogen}. The critical point for this model is occurred at \( p_c \approx 0.0715 \). Using the finite-size scaling relations articulated in Eqs.~\eqref{Eq:susc} - \eqref{Eq:Binder}, we determine that the optimal critical exponents for this model, which make the best collapse of all data for various \( N \), are consistent with those of the model in case (1). Specifically, the critical exponents are \( \gamma \approx 1.75 \), \( \beta \approx 0.125 \), and \( \nu \approx 1.0 \). These results indicate that cases (1) and (2) are identical, and both belong to the universality class of the two-dimensional Ising model \cite{stanley1971phase}. All the critical exponents of the model follow the identity relation \( \nu d_c = 2 \beta + \gamma \), where \( d_c = 2 \) is the critical dimension of the two-dimensional Ising model.

\section{Summary and outlook}
This paper studies the opinion dynamics of the Sznajd model on a complete graph and a two-dimensional square lattice. Each agent has two possible opinions, represented by the Ising number \( \pm 1 \), and these opinions are randomly assigned to the nodes of the graphs. The links, or edges, of the graphs, represent social connections within the social system. Agents in this model exhibit two types of behaviors: conformity (conformist agents) and independence (independent agents). Conformist agents follow the majority opinion in the population, while independent agents act independently, changing their opinions without being influenced by the group's opinion.

As stated in the original Sznajd model, two paired agents with the same opinion influence their neighbors to adopt their opinion. Expanding on the original model, we consider several paired agents of size \( r \) and influence their neighbors of size \( n \) for the model on the complete graph. For \( r = 2 \), the model reduces to the original Sznajd model on the complete graph. For \( n = 1 \), the model reduces to the \( q \)-voter model on the complete graph. In the two-dimensional square lattice, four paired agents influence eight of their neighbors whenever the paired agents have a unanimous opinion. If the four agents do not have a unanimous opinion, then two or three paired agents with unanimous opinions can still influence their neighbors to adopt their opinion. The neighboring agents act independently with probability \( p \), and with probability \( 1/2 \), they change their opinion \( \pm S_{i}(t) = \mp S_i(1+t) \). Otherwise, with probability \( 1 - p \), the neighboring agents follow the paired agent whenever there is an agreement among the paired agents.

For the model on the complete graph, we observe that the size of neighboring agents \( n \) does not influence the critical point at which the model undergoes an order-disorder phase transition. However, the evolution of the fraction opinion \( c \) converges to a steady state with distinct trajectories for different \( n \), following the relation \( t \sim 1/n \). The model exhibits a second-order (continuous) phase transition for \( r \leq 5 \) and a first-order (discontinuous) phase transition for \( r > 5 \) for all values \( n \neq 0 \). Utilizing finite-size scaling relations, we ascertain that the model on the complete graph belongs to the mean-field Ising universality class for all \( n \), characterized by critical exponents \( \beta \approx 0.5 \), \( \nu \approx 2.0 \), and \( \gamma \approx 1.0 \). Our analysis of the order-disorder phase transition, conducted through the effective potential and the stationary probability density function of the fraction opinion \( c \), yields consistent results.

For the model on the two-dimensional square lattice, both case (1) and case (2) exhibit a second-order phase transition. The critical points are identified as \( p_c \approx 0.0805 \) for case (1) and \( p_c \approx 0.0715 \) for case (2). However, based on finite-size scaling analysis, both cases share the same best-fit critical exponents: \( \beta \approx 0.125 \), \( \gamma \approx 1.75 \), and \( \nu \approx 1.0 \). The result indicates that both cases are essentially identical in their critical behavior. These critical exponents follow the identity relation \( \nu d_c = 2 \beta + \gamma \), where \( d_c = 2 \) is the critical dimension of the two-dimensional Ising model. The data suggest that the model for both cases belongs to the two-dimensional Ising universality class.

\section*{Data Availability}
Data will be made available on request.

\section*{CRediT authorship contribution statement}
\textbf{Azhari:} Conceptualization, Writing, Formal analysis, Review \& editing, Funding acquisition \& Supervision. \textbf{R.~Muslim:} Main Contributor, Methodology, Software, Formal analysis, Validation, Writing, Visualization, Review \& editing. \textbf{D. A. Mulya:} Simulation \& Visualization. \textbf {H. Indrayani:} Writing \& Visualization. \textbf {C. A. Wicaksana:} Writing \& Formal Analysis. \textbf{A. Rizki:} Formal analysis. All authors read and reviewed the paper.

\section*{Declaration of Interests}
The contributors declare that they have no apparent competing business or personal connections that might have appeared to have influenced the reported work.

\section*{Acknowledgments}
The authors would like to thank Kemendikbudristek (Ministry of Education, Culture, Research, and Technology of Indonesia) through the DRTPM-PKDN Scheme with contract number 69/UN5.2.3.1/PPM/KP-DRTPM/B/2023 for its financial support. Didi A. Mulya thanks BRIN talent management through the Research Assistant program with decree number 60/II/HK/2023.

\bibliographystyle{elsarticle-num}

\bibliography{cas-refs}

\vskip3pt
\end{document}